\pdfoutput=1
\documentclass[sigconf,nonacm]{acmart}
\acmConference[ESEM'24]{18th ACM/IEEE International Symposium on Empirical Software Engineering and Measurement}{Oct.'24}{Barcelona, Spain}

\newif\ifDEBUG
\DEBUGtrue

\newif\ifARXIV
\ARXIVfalse

\usepackage{threeparttable}
\usepackage{tabularx}
\usepackage{tabulary}
\usepackage{colortbl}
\usepackage{multicol}
\usepackage{multirow}
\usepackage{xurl}  

\usepackage{enumitem}
\usepackage{xspace}

\usepackage{pgfplots}
\pgfplotsset{compat=1.8}
\usepgfplotslibrary{statistics}

\usepackage{soul}

\ifDEBUG
    \newcommand{\JD}[1]{\textcolor{purple}{[JD says:#1]}}

\else
    \newcommand{\JD}[1]{}

\fi

\usepackage{tcolorbox}

\makeatletter
\def\cl@chapter{}
\makeatother
\usepackage{cleveref}

\crefformat{section}{\S#2#1#3}
\crefname{figure}{Figure}{Figures}
\crefname{appendix}{Appendix}{Appendices}
\crefname{table}{Table}{Tables}
\crefname{algorithm}{Algorithm}{Algorithms}
\crefname{listing}{Listing}{Listings}
\crefname{theorem}{Theorem}{Theorems}
\crefname{thm}{Theorem}{Theorems}
\crefname{lemma}{Lemma}{Lemmata}
\crefname{equation}{Eqt.}{Eqts.}
\crefformat{Grammar}{Grammar #1}

\newcommand{\ie}{\textit{i.e.,} }
\newcommand{\eg}{\textit{e.g.,} }
\newcommand{\etal}{\textit{et al.}\xspace}

\newcommand{\myparagraph}[1]{\textbf{#1}}

\newcommand{\code}[1]{{\small\texttt{#1}}\xspace}

\usepackage[font=small,skip=0pt]{caption}

\newcommand{\numPRs}{31,462\xspace}
\newcommand{\numGDPRPRs}{15,731\xspace}
\newcommand{\numRepos}{6,513\xspace}
\newcommand{\numSurvey}{56\xspace}

\newcommand{\MyTitle}[1]{}
\renewcommand{\MyTitle}{Understanding the Impact of Data Privacy Regulations on open-source Software}
\renewcommand{\MyTitle}{Legal compliance Meets Open-Source: A First Look in the Context of GDPR}
\renewcommand{\MyTitle}{An Exploratory Study of GDPR Compliance in Open-Source Projects}
\renewcommand{\MyTitle}{Perceptions of GDPR in Open-Source Software}
\renewcommand{\MyTitle}{Open-Source Perceptions of Privacy: A Case Study in GDPR Compliance}
\renewcommand{\MyTitle}{Perceptions of GDPR Compliance in Open-Source Software Development}
\renewcommand{\MyTitle}{Understanding the Impact of the General Data Protection Regulation (GDPR) on Open-Source Software Development}
\renewcommand{\MyTitle}{An Exploratory Mixed-Methods Study on General Data Protection Regulation (GDPR) Compliance in Open-Source Software}

\title{\MyTitle}

\author{Lucas Franke}
\email{lfranke@vt.edu}
\affiliation{%
  \institution{Virginia Tech}
  \city{Blacksburg}
  \state{Virginia}
  \country{USA}
}

\author{Huayu Liang}
\email{huayu98@vt.edu}
\affiliation{%
  \institution{Virginia Tech}
  \city{Blacksburg}
  \state{Virginia}
  \country{USA}
}

\author{Sahar Farzanehpour}
\email{saharfarza@vt.edu}
\affiliation{%
  \institution{Virginia Tech}
  \city{Blacksburg}
  \state{Virginia}
  \country{USA}
}

\author{Aaron Brantly}
\email{abrantly@vt.edu}
\affiliation{%
  \institution{Virginia Tech}
  \city{Blacksburg}
  \state{Virginia}
  \country{USA}
}

\author{James C. Davis}
\email{davisjam@purdue.edu}
\orcid{0000-0003-2495-686X}
\affiliation{%
  \institution{Purdue University}
  \city{West Lafayette}
  \state{Indiana}
  \country{USA}
}

\author{Chris Brown}
\email{dcbrown@vt.edu}
\affiliation{%
  \institution{Virginia Tech}
  \city{Blacksburg}
  \state{Virginia}
  \country{USA}
}

\begin{document}

\begin{abstract}


\textbf{Background:}
Governments worldwide are considering \textit{data privacy regulations}.
These laws, such as the European Union's General Data Protection Regulation (GDPR), require software developers to meet privacy-related requirements when interacting with users' data.
Prior research describes the impact of such laws on software development, but only for commercial software.
Although open-source software is commonly integrated into regulated software, and thus must be engineered or adapted for compliance, we do not know how such laws impact open-source software development.



\textbf{Aims:}
Understanding how data privacy laws affect open-source software development.
We focused on the European Union's GDPR, as it is the most prominent such law.
We specifically investigated 
  how GDPR compliance activities influence OSS developer activity (RQ1),
  how OSS developers perceive fulfilling GDPR requirements (RQ2),
  the most challenging GDPR requirements to implement (RQ3),
  and
  how OSS developers assess GDPR compliance (RQ4).

\textbf{Method:}
We distributed an online survey to explore perceptions of GDPR implementations from open-source developers (N=\numSurvey). To augment this analysis, we further conducted a repository mining study to analyze development metrics on pull requests (N=\numPRs) submitted to open-source GitHub repositories. 

\textbf{Results:} Our results suggest GDPR policies complicate open-source development processes and introduce challenges for developers, primarily regarding the management of users' data, implementation costs and time, and assessments of compliance. Moreover, we observed negative perceptions of GDPR from open-source developers and significant increases in development activity, in particular metrics related to coding and reviewing activity, on GitHub pull requests (PRs) related to GDPR compliance. 

\textbf{Conclusions:} Our findings provide future research directions and implications for improving data privacy policies, motivating the need for policy-related resources and automated tools to support data privacy regulation implementation and compliance efforts in open-source software.

\end{abstract}


\maketitle

\section{Introduction} \label{sec:Introduction}

\footnote{This work appeared at the 18th ACM/IEEE International Symposium on Empirical Software Engineering and Measurement (ESEM 2024).}
Software products collect an increasing amount of data from users to enhance user experiences through personalized, machine learning-enabled~\cite{pew2019} application behaviors~\cite{bucklin2009click} and marketing~\cite{hyper}.
Such practices may benefit users, but also threaten their well-being.
For example, in 2013, Facebook allowed the
  political research firm
Cambridge Analytica to access data on $\sim$87 million Facebook users~\cite{CambridgeFB}.
Cambridge Analytica used this data to influence US elections~\cite{nymag2019a,nymag2019b}.

To protect their citizens, over 100 governments worldwide are developing \emph{data privacy regulations}~\cite{unctad}.
Their goal is to constrain how their citizens' personal data is collected, processed, stored, and saved.
Some target specific industries, \eg the United States's Health Insurance Portability and Accountability Act (HIPAA), which places requirements on healthcare organizations handling medical data~\cite{hipaa1996}.
Others cover personal data regardless of context, \eg the European Union's General Data Protection Regulation (GDPR), which grants rights to EU citizens and affects entities that handle their data~\cite{gdpr2018}.
The penalties for non-compliance with data privacy laws and regulations may be severe~\cite{gdpr_83,akhlaghpour2021learning}.
For example, under GDPR, corporations have been fined millions or billions of euros~\cite{milmo2023facebook}.
\ul{Most organizations store and manipulate this data electronically through software, and so
ensuring the software is in legal compliance is an important software engineering task.}

Data privacy regulations create challenging software requirements because they entail both technical and legal expertise. Software developers must implement required features, such as obtaining consent from users for data collection, to ensure their organizations' products are compliant.
However, developers may have limited legal knowledge~\cite{verdon2006security,moquin2016roles} and receive minimal training~\cite{allan2007reskilling,holst2017liability}.
This can lead to coarse solutions, such as exiting the affected market~\cite{pwc2017} --- hundreds of websites simply banned all European users when GDPR went into effect~\cite{reutersinstitute2020,niemanlab2018}.
Researchers have explored the impact of data privacy regulations on businesses~\cite{li2019impact,layton2019social,pwc2017}, users~\cite{almeida2021exploring,bowyer2022human,kulyk2020has}, and observable software product properties such as website cookies~\cite{kretschmer2021cookie} and database performance~\cite{shastri_banakar_wasserman_kumar_chidambaram_2020}.
\ul{However, there has been limited study of how such laws affect the software development process.}
The few existing studies have been of commercial software development~\cite{alhazmi_arachchilage_2021,doi:10.1080/01972243.2019.1583296}; we lack knowledge of the effects of GDPR on open-source software (OSS) development. 

The goal of this work is to describe the impact of data privacy regulation compliance on open-source software. 
Our study is the first on this topic.\footnote{This paper is an extension on our preliminary work, presented as a poster~\cite{frankeICSEPoster2024}.}
We therefore adopt an exploratory methodology to provide an initial characterization and identify phenomena of interest for further study.
Our study draws on two data sources collected in two phases. 
The first phase examined qualitative data on developers' experiences with GDPR implementations in OSS, collected via a survey (N=\numSurvey). To further investigate the impact of GDPR in OSS, the second phase collected and analyzed developers' activities in open-source projects on GitHub, examining metrics and sentiments on \numPRs pull requests, divided into \numGDPRPRs GDPR and non-GDPR pull requests (PRs).

Our results show GDPR compliance negatively impacts open-source development---incurring complaints from developers and significantly increasing coding and reviewing activities on PRs. In addition, despite the benefits of data privacy regulations for users, we find developers have mostly negative perceptions of the GDPR, reporting challenges with implementing and verifying policy compliance. We also find that interactions with legal experts hinder development processes, yet developers rarely consult with legal teams---often relying on ad hoc methods to verify GDPR compliance. 

In sum, our contributions are:
\begin{itemize}[leftmargin=0.5cm,topsep=0pt]
    \item We survey OSS developers to understand developers' experiences with GDPR compliance and challenges with implementing and assessing data privacy regulations.
    \item We empirically analyze the impact of GDPR-related implementations on development activity metrics.
    \item We use natural language processing (NLP) techniques to evaluate the perceptions of GDPR compliance through discussions on OSS repositories.
\end{itemize}

\noindent
\ul{Significance:}
This work contributes an exploratory analysis on the impact of GDPR compliance on open-source software.
It identifies interesting phenomena for further research---in particular opportunities to support policy implementation and verification.
We also provide recommendations for policymakers and software developers to improve data privacy regulations and their implementation.

\ifARXIV
\else

\section{Background} 
\label{sec:Background}


\subsection{Software Regulatory Compliance} \label{sec:B-RegCompliance}

\subsubsection{In General}

Software requirements are divided into two categories: functional and non-functional~\cite{sommerville2011software}.
Functional requirements pertain to input/output characteristics, \ie the functions the software computes.
Non-functional requirements cover everything else, such as resource constraints, deployment conditions, and development process.
One major class of non-functional requirement is \emph{compliance with applicable standards and regulations}.
These requirements are typically developed and enforced on a per-industry basis in acknowledgment of that industry's risks and best practices~\cite{hobbs2019embedded}.

Complying with standards and regulations has been part of software engineering work for many years. 
%
Some standards apply to any manufacturing process, \eg the ISO 9001 quality standard~\cite{iso9001}.
Others are generic to software development (\eg ISO/IEC/IEEE 90003~\cite{iso90003}).
Still others are contextualized to the risk profile of the usage context,
  \eg
    ISO 26262~\cite{iso26262} or IEC 61508~\cite{iec61508} which describe standards for safety-critical systems~\cite{hobbs2019embedded};
    the US HIPAA law (Health Insurance Portability and Accountability Act) which describes privacy standards for handling medical data~\cite{hipaa1996};
    and
    the US FERPA law (Family Education Rights and Privacy Act) which describes privacy standards for handling educational data~\cite{ferpa1974}.
Although these regulations are not new (\eg FERPA dates to 1974, HIPAA to 1996, and IEC 61508 to 1998), software engineering teams still struggle to comply with them~\cite{lloyd2009iec,de2016industrial,carroll2016software,fischer2018critical}.

\subsubsection{In Open-Source Software}

This study focuses on GDPR compliance in open-source software.
The reader may be surprised that regulatory compliance is a factor in open-source software development, as open-source software licenses such as MIT~\cite{MITLicense}, Apache~\cite{apache2004}, and GNU GPL~\cite{gplv2} disclaim legal responsibility.
For example, the MIT license, the most common license on GitHub~\cite{github2015}, states \emph{``the software is provided `as is', without warranty...[authors are not] liable for any claim, damages, or other liability''}.
However, users and developers of open-source software may desire regulatory compliance.
We note three examples.
(1) A majority of open-source software is developed for commercial use~\cite{octoverse2022} and may require standards or regulatory compliance~\cite{vazao2019siem}.
(2) Users with open-source software components in software supply chains~\cite{harutyunyan2020managing,okafor2022sok} may request compliance requirements such as web cookies. The developers may service these requests.
(3) Users may extend open-source software themselves and undertake their own compliance analysis~\cite{STOKES2012481}.
Standards such as IEC 61508--Part 3 include provisions for doing so~\cite{IEC61508-3}. 

Open-source software is no longer a minor player in commercial software engineering.
Multiple estimates suggest that open-source components comprise the \emph{majority} of many software applications~\cite{octoverse2022,nagle2022census}.
In a 2023 survey of $\sim$1700 codebases across 17 industries, Synopsys found open-source software in 96\% of the codebases and reported an average contribution of 75\% of the code in the codebase~\cite{synopsys2023report}.
It is therefore important to understand how open-source software development considers non-functional requirements such as regulatory compliance.

\subsection{Privacy Regulations, Especially GDPR} \label{sec:B-PrivacyReg} 

\subsubsection{Consumer Privacy Laws}

In~\cref{sec:B-RegCompliance} we discussed standards and regulatory requirements that affect software products based on industry.
Recently a new kind of regulation has begun to affect software: consumer privacy laws. 
The most prominent example of such a law is the European Union's General Data Protection Regulation (EU GDPR), enacted in 2016 and enforceable beginning in 2018.
Examples in the United States include 
  the California Consumer Privacy Act (CCPA, enacted 2018) 
  and
  the Virginia Consumer Data Protection Act (CDPA, enacted 2021). 
Similar legislation has been considered by $>$100 governments~\cite{unctad,iapp_privacy_chart}.

\subsubsection{The General Data Protection Regulation (GDPR)}

The General Data Protection Regulation (GDPR)~\cite{gdpr2018} protects the personal data of European Union (EU) citizens, regardless of whether data collection and processing is based in the EU.
The law has implications for entities that interact with the personal data of EU citizens, divided into
  data \textit{subjects},
  data \textit{controllers},
  and
  data \textit{processors}~\cite{gdpr_4}.
Data subjects are individuals whose personal data is collected.
Data controllers are any entities ---organization, company, individual, or otherwise --- that own, control, or are responsible for personal data.
Data processors are entities that process data for data controllers. 
The GDPR grants data subjects rights to their personal data, providing guidelines and requirements to data controllers and processors to understand how to properly handle this data.

GDPR compliance is complex for software engineers and consequential for their organizations.
Data controllers and processors commonly use software, \eg a controller's mobile app transmits data to its backend service and processors subsequently access and update the database.
Software teams must determine appropriate data policies, update their systems to comply, and validate them, \eg incorporating cookie consent notices into websites to provide users with informed consent~\cite{utz2019informed}.
Anticipating a lengthy compliance process, the EU enacted the GDPR in 2016 but made it enforceable in 2018, allowing two years for corporations to prepare~\cite{European_Data_Protection_Supervisor}.
Companies in the US and UK alone invested \$9 billion in GDPR compliance~\cite{Vuleta_2023}.
As of December 2022, many use manual compliance methods or are not compliant~\cite{cytrio2023}.
Non-compliance is costly: 
  thousands of distinct fines have been imposed on non-compliant data controllers and processors, exceeding {\texteuro}2.5 billion~\cite{enforcementtracker2023}.   

Although GDPR compliance affects any software that processes the data of EU citizens, and open-source software components comprise the majority of many software applications that process such data~\cite{octoverse2022,synopsys2023report,nagle2022census}, \textit{to the best of our knowledge there is no prior research on the impacts of GDPR compliance in open-source software}.

\section{Methodology}

\subsection{Data Availability and Research Questions}

In~\cref{sec:Background} we described a range of privacy-related standards and regulations.
We noted that there has been little study of the effect of these requirements on open-source software engineering practice.
To address this gap, we need data.
\cref{table:DataAvailabilityByPrivacyLegislation} estimates the availability of software engineering data associated with these requirements through two common metrics:
  the number of posts on Stack Overflow
  and
  the number of pull requests on GitHub.


\begin{table}[h!]
\small
\centering
\caption{
  Software engineering data availability for privacy legislation.
  Data from keyword search on Nov. 13, 2023.
  We studied GDPR.
  }
\label{table:DataAvailabilityByPrivacyLegislation}
\begin{tabular}{lccc}
\toprule
\textbf{Privacy Law (Year)}   & \textbf{Stack Overflow} & \textbf{GitHub-PRs} \\
\midrule
GDPR (2016)        & 2058           & 64\ K       \\
\midrule
HIPAA (1996)       & 725            & 5\ K        \\
CCPA (2018)        & 96             & 1\ K        \\
FERPA (1974)       & 35             & 254       \\
CDPA (2021)        & 7              & 19        \\
PIPEDA (2000)      & 5              & 31        \\
\bottomrule
\end{tabular}
\end{table}

Based on this data, we scoped our study to the EU's GDPR; and to open-source software hosted on GitHub, currently the most popular hosting platform for OSS.
We answer four research questions:

\begin{itemize}
    \item[\textbf{RQ1:}] How does GDPR compliance influence development activity on OSS projects? 
    \item[\textbf{RQ2:}] How do OSS developers perceive fulfilling GDPR requirements?
    \item[\textbf{RQ3:}] What GDPR concepts do OSS developers find most challenging to implement?
     \item[\textbf{RQ4:}] How do OSS developers assess GDPR compliance?
\end{itemize}







We analyzed data from quantitative and qualitative sources:
  surveying open-source developers and mining OSS repositories on GitHub.
We present how we obtained and analyzed each data source next.
We integrate this data in answering RQ1 and RQ2, and use the survey data alone to answer RQ3 and RQ4. 

\subsection{Data Source 1: Developer Survey}


To explore the impact of implementing GDPR policies on OSS development, we distributed an online survey for open-source developers. 
This data informed our answers to all RQs.
We used a four-step approach motivated by the framework analysis methodology~\cite{frameworkanalysis} for policy research to collect and analyze data in the second phase of our experiment.
An overview of this process is presented in~\cref{tab:phase2}.
Our \textbf{Institutional Review Board (IRB)} provided oversight.

{
\begin{table*}[tbp]
    \centering
    \caption{
    Overview of sample questions from pilot interview study and survey design/analysis for framework analysis approach used for Data Source 2.
    The final column notes the inter-rater agreement score for these themes using the $\kappa$ score, prior to reaching agreement.
    }
\small
    \begin{tabular}{cc|ccc}
        \toprule
        \textbf{Interview Question} & \textbf{Codes} & \textbf{Survey Question} & \textbf{Codes} & \textbf{$\kappa$} \\
        \toprule
        \multirow{4}{0.28\linewidth}{What meaningful impact, if any, do you believe the GDPR has had on data security and privacy?} & \multirow{4}{0.12\linewidth}{data privacy, rights to users, data collection} & \multirow{4}{0.28\linewidth}{What impact, if any, do you believe the GDPR and similar data privacy regulations have had on data security and privacy?} & \multirow{4}{0.2\linewidth}{\small data privacy, data processing, data collection, insufficient information, data breach, fines} &  \\
        & & & & \\
        & & & & \\
        & & & & 0.736 \\ \midrule
        \multirow{7}{0.280\linewidth}{What GDPR concepts do you find the most difficult or frustrating to implement?} & \multirow{7}{0.12\linewidth}{None, data minimization, embedded content} & \multirow{7}{0.28\linewidth}{What GDPR concepts do you find the most difficult or frustrating to implement?} & \multirow{7}{0.2\linewidth}{\small privacy by design, data minimization, cost, data processing, user experience, data management, security risks, None, lawfulness and dispute resolution, time, right to erasure} &  \\
        & & & & \\
        & & & & \\
        & & & & \\
        & & & & \\
        & & & & \\
        & & & & 0.929 \\ \midrule
        \multirow{4}{0.280\linewidth}{Have you had to specifically seek out legal consultation on GDPR-related issues, and if so, how did that affect your development process?} & \multirow{4}{0.12\linewidth}{Yes/No; no effect, negative effect (time)} & \multirow{4}{0.280\linewidth}{Have you had to specifically seek out legal consultation on GDPR-related issues, and if so, how did that affect your development process?} & \multirow{4}{0.2\linewidth}{\small Yes/No; N/A, no effect, positive effect, negative effect (cost, time, data storage, data processing,...)} & \\
        & & & & \\
         & & & & \\
        & & & & 0.514 \\ \midrule
        \multirow{6}{0.280\linewidth}{During your software development projects, do you frequently consult with a legal team, and if so, how does this impact the development processes? If not, how did you assess GDPR compliance for your software projects?} & \multirow{6}{0.12\linewidth}{Yes: legal consultation; No: privacy by design, data minimization} & \multirow{6}{0.280\linewidth}{During your software development projects, have you consulted with a legal team? If not, how do you assess GDPR compliance for your software projects?} & \multirow{6}{0.2\linewidth}{Yes: legal consultation; No: accountability system, online resources, self-assessment, data management, none), N/A} & \\
        & & & & \\
        & & & & \\
        & & & & \\
        & & & & \\
        & & & & 0.668 \\ \midrule
         \multirow{4}{0.280\linewidth}{---} &\multirow{3}{0.12\linewidth}{---} & \multirow{4}{0.280\linewidth}{Has implementing GDPR concepts for compliance impacted your development process in any way? (\textit{yes/no/maybe}) \newline Please explain:} & \multirow{3}{0.2\linewidth}{\small positive impact (logging, privacy by design), negative impact (cost, data management, security,...), no impact } \\
        & & & & \\
        & & & & \\
        & & & & 0.860 \\
        \bottomrule
    \end{tabular}
    \label{tab:phase2}
\end{table*}
}    

\subsubsection{Step 1: Pilot Study and Data Familiarization}
To formulate an initial thematic framework for our qualitative analysis, we conducted semi-structured pilot interviews with OSS developers ($n = 3$).
As no prior work has explored the perceptions of GDPR compliance in OSS, pilot interviews gave us insight into developers' perceptions and experiences with implementing GDPR concepts in the context of open-source software development. 
Two subjects had contributed to PRs in our dataset, and the third was a personal contact.
They had a wide range of open-source development experience, from $<1$ year to $>20$ years.
Interviews were transcribed using Otter.ai 
and coded by two researchers to inform our survey.

Thematic analysis of our pilot interviews provided insight that informed our survey questions. 
The participants highlighted the challenges with implementing GDPR requirements in open-source software.
One participant worked at a large corporation and outlined differences between GDPR compliance at their company and in OSS, namely with (1) approaches used to assess whether compliance is implemented correctly, and (2) access to legal teams.
The other two participants discussed the impact of the GDPR, noting its privacy benefits as well as challenges OSS developers face implementing GDPR requirements and assessing compliance.
These findings informed our survey. 

 
\subsubsection{Step 2: Survey Design}

The survey consisted of open-ended and short answer questions seeking details about GDPR implementation and experiences in the context of open-source software development.
We used the pilot study interview results to identify topics to focus on in the survey.
Based on the interviews, we asked about the perceived impact of the GDPR on data privacy, the most difficult concepts to implement, and how they assess GDPR compliance.
The survey instrument is in the supplemental material.

\subsubsection{Step 3: Participant Recruitment}

We distributed our survey in three rounds.
In the first round, we emailed a sample of 98 developers who authored or commented on GDPR-related pull requests with a publicly available email addresses.
We received 5 responses, \ie a 5\% response rate.
In the second round, we made broader calls for participation on Twitter and Reddit.
We received 44 responses, 2 of which indicated no experience implementing GDPR compliance. All survey respondents in these rounds were entered in a drawing for two \$100 Amazon gift cards.
After a few months, we undertook the third round, redistributing our survey to an additional 235 GitHub users with GDPR implementation experience (authored GDPR-related pull requests in our dataset) and offered individual compensation (\$10 gift card) to encourage participation.
We received 9 responses (4\% response rate).
In total we have data from \numSurvey survey participants (14 from direct GitHub contacts and 42 from Twitter and Reddit).

Our participants have a median of approximately 5 years of OSS development experience (avg = 5.9) and 6 years of general industry experience (avg = 7.7).
Participants reported contributing to a variety of OSS projects such as Mozilla, Wordpress, Fedora, Moodle, Ansible, Flask, Django, Kubernetes, PostGreSQL, OpenCV, GitLab, and Microsoft Cognitive Toolkit.


\subsubsection{Step 4: Data Analysis}

To analyze our survey results, we used an open coding approach.
Two researchers independently performed a manual inspection of responses--highlighting keywords and categorizing responses based on the pre-defined themes derived from our pilot study.
If new themes arose, the coders discussed and agreed upon adding the new theme.
Then, both coders came together to merge their individual results. Finally, we used Cohen's kappa ($\kappa$) to calculate inter-rater agreement (see Table~\ref{tab:phase2}).

\subsection{Data Source 2: GDPR PRs on GitHub}

We collected data concerning GDPR compliance by analyzing pull requests on GitHub repositories. 
Pull requests are a mechanism on GitHub that allow developers to collaborate on open-source repositories, involving code contributions from developers to be reviewed and merged into the source code~\cite{github2023pull}.

\subsubsection{GDPR and non-GDPR PRs}\label{sec:prs}
We used the GitHub REST API to search for \textit{GDPR-related pull requests}---pull requests returned by the GitHub API's default search with the query string ``GDPR''.
Manual inspection suggested the results are typically English-language PRs related to (GDPR) data privacy regulatory compliance.

Using this method, we collected GDPR-related PRs created from April 2016 (when the GDPR was adopted by the European Parliament) to January 2024. We removed content submitted by users with ``bot'' in their username~\cite{abdellatif2022bothunter} and designated as a bot type according to the GitHub API\footnote{\url{https://docs.github.com/en/graphql/reference/objects\#bot}} to avoid PRs generated by automated systems.
This resulted in \numGDPRPRs GDPR-related pull requests across \numRepos unique GitHub repositories.
For comparison, we also collected a random sample of \numGDPRPRs pull requests created in these same repositories after April 2016 that did \textbf{not} mention ``GDPR'', which we call \textit{non-GDPR-related pull requests}.
The studied repositories had a median of 14 stars (avg = 1,635), 11 forks (avg = 416),
  727 commits (avg = 8,997),
  172 PRs (avg = 1,425), and 15 contributors (avg = 59),
suggesting popular, active repositories.
The distribution of PRs across all repositories in our GDPR-related and non-GDPR-related datasets is summarized in Table~\ref{tab:PRs}.


\begin{table}[tbp]
    \centering
        \caption{
        Distribution of PRs in Datasets.
        }

    \begin{tabular}{lrrrrr}\hline
       \textbf{Dataset}  & \textbf{min} & \textbf{50\%ile} & \textbf{75\%ile} & \textbf{90\%ile} & \textbf{max} \\ \hline
        GDPR & 1 & 1 & 2 & 3 & 956 \\ \hline
        non-GDPR & 1 & 2 & 10 & 34 & 203 \\ \hline
    \end{tabular}
    \label{tab:PRs}
\end{table}

\subsubsection{Measuring Development Activity}
To analyze GDPR's impacts, we collected development activity metrics~\cite{pullreqs} per pull request:
\begin{itemize}[topsep=0pt]
    \item \textit{Comments:} the total number of comments
    \item \textit{Active time:} the amount of time the PR remained active (until merged or closed)
    \item \textit{Commits:} the total number of commits
    \item \textit{Additions:} the number of lines of code added
    \item \textit{Deletions:} the number of lines of code removed
    \item \textit{Changed files:} the total number of modified files
    \item \textit{Status:} outcome of PR (merged, closed, or open)
\end{itemize}


We selected these metrics to analyze development activity, specifically to derive coding and code review tasks from pull requests. We compared the distributions of these metrics between GDPR-related and non-GDPR-related PRs using a Mann-Whitney U test, to compare nonparametric ordinal data between the datasets~\cite{macfarland2016mann}. To control for multiple comparisons on the same dataset, we calculate adjusted p-values using Benjamini-Hochberg correction~\cite{888cd474-50a6-33fd-a789-415b80e67e78}. We measure effect size ($r$) for significant results using Cohen's $d$~\cite{cohen2013statistical}. 


\subsubsection{Measuring Developer Perception}
To augment our survey results, we applied sentiment analysis---a technique to automatically infer sentiment from natural language---on the title, body, commit messages, review comments, and discussion comments from pull requests in our datasets to examine developer perceptions of GDPR compliance. Prior studies have similarly inferred developer sentiment and emotion from GitHub activity, including PR discussion comments~\cite{pletea2014security}, review comments~\cite{huq2019understanding}, commit messages~\cite{guzman2014sentiment}, and bodies~\cite{park2021assessing}.
While this technique sometimes has negative results in software engineering contexts~\cite{jongeling2017negative}, we use it in our exploratory work as a proxy to obtain preliminary insights into developers' sentiments regarding GDPR compliance in OSS. 

We followed standard NLP preprocessing steps~\cite{kumar2022sentiment}:
(1) We removed bot-generated content using the process described in Section~\ref{sec:prs}.
(2) We removed non-sentiment material: hyperlinks and mentions (``@username''). 
(3) We tokenized text
using the Natural Language Toolkit (NLTK) \code{tokenize} library. 
(4) We converted tokens to lowercase and removed punctuation.
(5) We removed stopwords such as ``but'' and ``or'' (\code{nltk.corpus} library). 
(6) We lemmatized the text, \ie reducing words to their base form (\eg ``mice'' becomes ``mouse''~\cite{anandarajan2019text}) using \code{WordNetLemmatizer} from the \code{nltk.stem} library.
(7) We normalize the data by removing meaningless tokens, such as SHA or hash values for commits, and non-standard English words, such as words that contain numerical values (\ie ``3d'')~\cite{sproat2001normalization}.

After preprocessing the data, we were left with 15,731 titles, 14,515 bodies, 15,217 commit messages, 4,922 review comments, and 4,862 discussion comments across the GDPR-related pull requests. We compared these against non-GDPR-related PRs, for which we had 15,731 titles, 13,718 bodies, 15,652 commit messages, 3,427 review comments, and 3,165 discussion comments.

To perform sentiment analysis, we use three state-of-the-art models: Liu-Hu~\cite{hu2004mining}, VADER~\cite{hutto2014vader}, and SentiArt~\cite{jacobs2019sentiment}.
We fed the preprocessed textual data to each model, which provided compound sentiment scores. We use a t-test ($t$) to statistically analyze sentiment across our datasets. Moreover, we aim to assess the impact of the GDPR on developer sentiment over time. To accomplish this, we divided the GDPR and non-GDPR PRs into 3-month segments based on the creation date of the PR. Then, we performed sentiment analysis on the binned data to observe whether and how developer sentiments manifest in OSS interactions over the lifecycle of the GDPR regulation --- from its initial adoption in 2016, enforcement in 2018, and to the present. We combined all preprocessed textual elements (title, body, commit messages, review comments, and discussion comments) to observe the overall trends in PR communications and compare with non-GDPR data as a baseline sentiment in developer communications for the projects studied.

\section{Results}


We are interested in understanding the impact of GDPR implementations on open-source software by analyzing development activity and developer perceptions, including challenges with implementation and assessment of compliance. In this work, we answer our research questions using multiple sources---analyzing GitHub repositories and surveying open-source developers.
For RQ1 and RQ2, we report views from the survey and the GitHub measurements.
For RQ3 and RQ4, we use data only from the survey.

\subsection{RQ1: Development Activity}\label{sec:rq1}

This question was: \textit{RQ1: How does GDPR compliance influence development activity on OSS projects?}

\subsubsection{Survey} \label{sec:survey}

We surveyed \numSurvey OSS developers to understand the impact of GDPR implementations on development activity.
Most participants ($n = 41$, 73\%) responded ``Yes'' to a question regarding the impact of implementing GDPR concepts on development processes, indicating data privacy compliance effects open-source development.
When asked to elaborate, 23 developers provided examples of development impacts related to the GDPR.

\myparagraph{Data Management:} 11 participants mentioned GDPR requirements related to data management impact development activity, notably increasing development efforts. For instance, responses indicated handling personal data (P17) and anonymization (P19), managing data controllers (P21) and data recipients (P23), implementing functionality to limit the collection of personal data (P26), and the monitoring of data subjects from the EU (P28) all impacted development processes.  P53 also added ``\textit{we had to separate in a clear way sensitive data from the other data}'', exemplifying the effort needed to implement compliant data processing in OSS.

\myparagraph{Time and Costs:} Five participants mentioned GDPR compliance increases development time and costs in OSS. For example, regarding time, respondents said ``\textit{it does slow down our development cycle}'' (P54) and ``\textit{we lost a complete year to be ready}'' (P56). For costs, participants said ``\textit{budgets have soared}'' (P5) and ``\textit{costs of production should not go over the cost of consequence of data breach}'' (P46).

\myparagraph{Design:} Three participants also noted the effects of GDPR compliance on the design and structure of software products. For example, P54 responded ``\textit{we have to check whether we comply with GDPR every time we draft a new design}'' and P55 added ``\textit{the design of systems now incorporates the concept of needing to remove PII after the fact}''. P21 explained how GDPR compliance reduced the quality of their application's design--replying ``\textit{the principle of minimum scope was not observed}''---indicating potential unnecessarily extended scopes of variables in the code~\cite{chisnall2012go}.

\myparagraph{Organization:} Three participant responses embodied the negative effects of data privacy regulations on their organization, stating the GDPR has a ``\textit{major impact}'' requiring ``\textit{an overhaul of project management and program priorities}'' (P1). P45 highlighted that ``\textit{making sure to follow privacy by design}'' is challenging for GDPR compliance in OSS development. One participant also mentioned additional steps to verify implementations affected their development, stating ``\textit{we need to make an additional review with the GDPR consultants that functionality that is related to the users' data}'' (P53).

\myparagraph{Benefits:} One participant mentioned benefits to their development team and processes regarding the implementation of GDPR concepts, stating it helped highlight ``\textit{things we had not considered before}'', such as ensuring that ``\textit{logging functionality}'' and ``\textit{access restrictions}'' were in place (P1). However, the majority of responses indicate that GDPR compliance often increases development efforts and incurs negative impacts for open-source developers.
  
\subsubsection{Pull Request Metrics} \label{sec:rq1_prs}

To further observe the impact of GDPR compliance on OSS, we compared metrics for GDPR and non-GDPR related PRs.
\cref{tab:rq1_metrics} presents these results.
Using a Mann-Whitney U test, we found statistically significant differences between GDPR and non-GDPR PRs in the number of comments, active time, number of commits, lines of code added, lines of code deleted, and number of modified files. We also calculate the effect size for these results.

\textbf{This indicates that incorporating changes related to the GDPR has a major impact on development work}, leading to:
  increased discussions between developers, 
  longer review times,
  more code commits,
  and higher code churn. While we observed significant differences exist in pull request metrics between GDPR and non-GDPR PRs, the calculated effect sizes are ``small''~\cite{lakens2013calculating}, indicating low practical differences between the groups. Yet, these findings support our survey results from open-source developers purporting that GDPR compliance efforts affect OSS development. 


\begin{tcolorbox} [width=\linewidth, colback=yellow!30!white, top=1pt, bottom=1pt, left=2pt, right=2pt]
\textbf{Finding 1:} Developers report implementing GDPR compliance negatively affects development processes--citing cost, time, and data management as concerns. 

\textbf{Finding 2:} PRs related to GDPR compliance have significantly more development activity for coding (\textit{commits, additions, deletions, files changed}) and review (\textit{comments, active time}) tasks. 

\end{tcolorbox}


 


{
\begin{table}[th]
\small
\centering
\caption{
GDPR (G) vs. Non-GDPR (non-G) GitHub Activity Metrics.
}
\begin{threeparttable}
\begin{tabular}{|l|l|r|r|c|} \hline
  \textbf{Characteristic} & \textbf{Type} & \textbf{Median} & \textbf{\textit{p}-value} \\ \hline
 \multirow{2}{*}{\em Comments*} & G &  1  &   $p < 0.0001$  \\ 
 & non-G & 1 & ($U = 1.4E8$, $r = 0.09$) \\
 \hline
 \multirow{2}{*}{\em Active time (days)*}
 & G & $418.05$ &  $p < 0.0001$  \\
 & non-G & $1.78$ & ($U = 1.4E8$, $r = 0.14$) \\
  \hline 
 \multirow{2}{*}{\em Commits*} 
 & G  & 2 & $p < 0.0001$   \\
 & non-G  &  1 & ($U = 1.4E8$, $r = 0.04$) \\
 \hline 
 \multirow{2}{*}{\em Additions*} 
 & G & 57 & $p < 0.0001$  \\ 
 & non-G & 19  & ($U = 1.5E8$, $r = 0.05$) \\
 \hline 
  \multirow{2}{*}{\em Deletions*} 
 & G &  7 & $p < 0.0001$  \\ 
 & non-G  & 4 & ($U = 1.3E8$, $r = 0.05$) \\\hline
 \multirow{2}{*}{\em Changed files*} 
 & G &  4 & $p < 0.0001$  \\ 
 & non-G &  2 & ($U = 1.4E8, r = 0.03$) \\\hline
\end{tabular}
\begin{tablenotes}
\centering
\item \textbf{*} denotes statistically significant results (\textbf{p-value < 0.05})
\end{tablenotes} 
\end{threeparttable}
\label{tab:rq1_metrics}
\end{table}
}


\subsection{RQ2: GDPR Perceptions}\label{sec:rq2}

This question was: \textit{RQ2: How do OSS developers perceive fulfilling GDPR requirements?}


\subsubsection{Survey}

We asked participants their perceptions on the impact of GDPR regulations on privacy. 
Of participants who responded to this question ($n = 25$), most had negative opinions of the GDPR. 
Three participants were neutral (\eg ``\textit{N/A}'' (P4)).
We summarize positive and negative perceptions next. 

\myparagraph{Negative Perceptions:} Despite the utility of data privacy regulations, 22 participants reported negative perceptions of the GDPR. These responses primarily focused on three issues: cost, organizations, and enforcement.
For costs, respondents noted that implementing GDPR requirements is expensive and burdensome.
Participants said that compliance is ``\textit{costly for many companies}'' (P16) is ``\textit{too expensive}'' (P24), and ``\textit{the cost of protection should not go over the cost of consequence of data breach...GDPR [isn't] worth the time}'' (P46). P55 also highlights that ``\textit{in general there have been major costs to companies of all sizes}'' regarding GDPR implementations.
For organizations, participants reported a negative impact of the GDPR on companies and organizations.
They mentioned that GDPR compliance ``\textit{weakens small and medium-sized enterprises}'' (P15), ``\textit{threatens innovation}'' (P18), ``\textit{fails to meaningfully integrate the role of privacy-enhancing innovation and consumer education in data protection}'' (P23), and that ``\textit{in order to be safer than risky useful functionality is removed}'' (P52).
P46 added that the GDPR is ``\textit{a lot of headache...jobs for lawyers at the expense of people who are trying to solve real problems}''.
For enforcement, one subject said ``\textit{there is a large gap in GDPR enforcement among member states} (P17) and another observed ``\textit{the trend...is an increase in the number of times and the amount of fines}'' (P18). Similarly, P49 described GDPR as ``\textit{a big hammer}'', but was unsure ``\textit{if it has necessarily increased security and privacy at this point}''.

\myparagraph{Positive Perceptions:}
Eight participants had positive perceptions of the GDPR, generally stating that GDPR enhances data privacy for users.
For example, participants said that
  ``\textit{the risk of incurring and paying out hefty fines has made companies take privacy and security more proactively}'' (P30),
  that GDPR brings ``\textit{awareness to the importance about privacy}'' (P45),
  that
  ``\textit{data integrity is ensured}'' (P47), and ``\textit{customers can now delete their data quite easily}'' (P54). Participants also appreciated the increased accountability for corporations in safeguarding users' data---for example one participant stated ``\textit{Before GDPR data protection was usually considered only as an afterthought if not an outright joke. Nowadays companies will at least consider what they are doing wrong before violating data protection laws, rather than doing it by accident because no-one even thought about it}'' (P50).
  These responses reflect the intentions of the GDPR ---
  to safeguard the rights of users and their data online.

\subsubsection{Sentiment Analysis}\label{sec:rq2_sent}

We investigated the sentiment of developers implementing GDPR concepts by analyzing PR titles, commit messages, review comments, discussion comments, and bodies. Our overall results are in~\cref{tab:rq2_sentiment}. We anticipated a higher percentage of negative comments for GDPR-related pull requests. However, we did not find evidence that GDPR-related PRs have less favorable sentiments from developers. In fact, we found they often had \textit{more} positive sentiments than non-GDPR-related PRs---with two of the three models (Liu-Hu and VADER) indicating a statistically significant difference between the GDPR and non-GDPR sentiment. We speculate two explanations. First, non-GDPR-related PRs represent a broad range of code contributions, which could address a number of issues. Second, we are limited by the capabilities of the sentiment analyzer. For example, the two most negative commit messages for non-GDPR pull requests said ``\textit{obsolete}'' and ``\textit{fatal}'', which are common terms of art in software maintenance tasks~\cite{wnuk2013obsolete,rinard2007automated} (\eg ``\textit{fix fatal error}''). We also observed some variation at the beginning and end of our dataset collection period, but no significant variation in sentiment over time (see~\cref{fig:sentiment}).

\begin{table}[tbp]
    \centering
    \small
    \caption{
    GDPR (G) vs Non-GDPR (non-G) Sentiment Analysis
    }
    \begin{threeparttable}
    \begin{tabular}{|l|r|r|r|c|}
    \hline
       \textbf{Test} & \textbf{Type} & \textbf{Mean} & \textbf{Variance}  & \textbf{\em p-value} \\ \hline
       \multirow{2}{*}{\textit{Liu-Hu*}} & G & 0.43 & 0.27 & $p < 0.0001$ \\ 
        & non-G & -0.04 & 0.28 & ($t = -4.05$, $r = 0.22$) \\ \hline
        \multirow{2}{*}{\textit{VADER*}} & G & 0.44  & 0.04 & $p < 0.0001$ \\ 
        & non-G & 0.21 & 0.01 & ($t = -6.47$, $r = 0.02$) \\ \hline
        \multirow{2}{*}{\textit{SentiArt}} & G &  0.39 & 0.01 & $p = 0.1399$ \\ 
        & non-G & 0.36 & 0.002 & ($t = -1.10$, $r = 0.01$) \\ \hline
    \end{tabular}
    \begin{tablenotes}
    \centering
    \item \textbf{*} denotes statistically significant results (\textbf{p-value < 0.05})
    \end{tablenotes} 
    \end{threeparttable}
    \label{tab:rq2_sentiment}
\end{table}

\begin{figure}[tbp]
    \centering
    \begin{tikzpicture}
\begin{axis}[
    xlabel={Year},
    ylabel={Mean Sentiment},
    xmin=0, xmax=32,
    ymin=-2, ymax=2,
    xtick={0,4,8,12,16,20,24,28,32},
    xticklabels = {2016, 2017, 2018, 2019, 2020, 2021, 2022, 2023, 2024},
    ytick={-2,-1.5,-1,-0.5,0,0.5,1,1.5,2},
    legend pos=north west,
    ymajorgrids=true,
    grid style=dashed,
]
\addplot[color=black,mark=square] table[x index=0,y index=1] {figdata.txt};\label{p1} 
\addplot[color=red, mark=square] table[x index=0,y index=2] {figdata.txt};\label{p2} 
\addplot[color=blue, mark=square] table[x index=0,y index=3] {figdata.txt};\label{p3} 
\addplot[color=black,mark=*] table[x index=0,y index=4] {figdata.txt};\label{p4} 
\addplot[color=red, mark=*] table[x index=0,y index=5] {figdata.txt};\label{p5} 
\addplot[color=blue, mark=*] table[x index=0,y index=6] {figdata.txt};\label{p6} 

    
\end{axis}
\node [draw,fill=white] at (rel axis cs: 0.75,0.2) {\shortstack[l]{
\footnotesize\ref{p4} Liu-Hu (G) \\
\footnotesize\ref{p5} VADER (G) \\
\footnotesize\ref{p6} SentiArt (G) \\
\footnotesize\ref{p1} Liu-Hu (non-G)\\
\footnotesize\ref{p2} VADER (non-G) \\
\footnotesize\ref{p3} SentiArt (non-G)}};
\end{tikzpicture}
    \caption{
    Longitudinal GDPR (G) and Non-GDPR (non-G) Sentiment Analysis Data.
    We grouped GDPR and non-GDPR data into 3-month segments and used 3 sentiment models. 
    For each model, GDPR data is plotted in a color with a filled marker, and non-GDPR data in the same color but with a hollow marker.
    The general trend is that sentiment for GDPR data is moderately positive, and more positive than for non-GDPR data.
    }
    \label{fig:sentiment}
    \Description{Longitudinal sentiment analysis}{}
\end{figure}
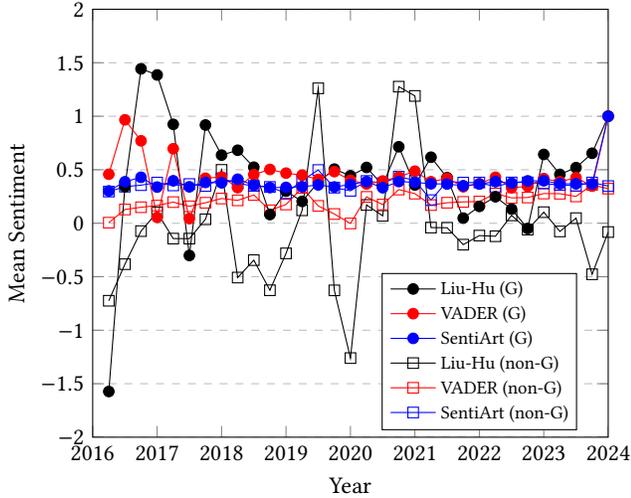

Nonetheless, manual inspection of negatively scored content showed OSS developers expressing frustration with GDPR compliance. For instance, one title and commit message described GDPR-related changes to ``\textit{avoid lawsuits by mentioning cookies thing}''~\cite{sethi2021pr}. Another title states adding ``\textit{just enough EULA [end user license agreement] not to get banned}''~\cite{betts2019pr}. Similar frustrations were shared in a PR body for ``\textit{GDPR stuff}'' adding changes to ``\textit{display the annoying cookies banner}''~\cite{tubin2023pr}. Discussion comments, such as ``\textit{will this conflict with GDPR?}''~\cite{llun2018comment}, also highlight OSS developers' confusion with GDPR requirements.

\begin{tcolorbox} [width=\linewidth, colback=yellow!30!white, top=1pt, bottom=1pt, left=2pt, right=2pt]
\textbf{Finding 3:} Despite its nominal advantages, most developers had \textit{negative} perceptions of the GDPR and its implementation.

\textbf{Finding 4:} We found developers did \textit{not} express more negative sentiments about GDPR compliance in PR discussions.

\textbf{Finding 5:} Sentiment related to GDPR compliance appears to be stable over time.

\end{tcolorbox}

\subsection{RQ3: Implementation Challenges}\label{sec:rq3}

This question was: \textit{RQ3: What GDPR concepts do OSS developers find most challenging to implement?}
In the survey data, we observed three common challenges: data management, data protection, and vague requirements.

\myparagraph{Data Management:} 11 developers responded that processing and storing users' data according to GDPR requirements is the most challenging concept to implement. For example, participants mentioned challenges implementing ``\textit{data protection}'' (P24), handling ``\textit{personal data}'' (P34), the ``\textit{exchange of documents containing personal data}'' (P32), the ``\textit{improper storage}'' (30) of user data, and ``\textit{knowing what info can or cannot be accessed or saved}'' (P49). In particular, four participants mentioned users' right to erasure---or the obligation for data controllers to delete users' data upon request ``without undue delay''~\cite{erasure}---as the most complicated requirement to implement. For example, P53 responded, ``\textit{it's not always easy enough to implement data processing in a way, that it's anonymized, and if the user would like their data to be erased, be able to continue processing of the results based on user data in an anonymous way}''--- describing the complexity of this requirement for their project.

\myparagraph{Data Protection:} Five participants mentioned security factors as a challenge for GDPR compliance.
For instance, participants were concerned with ``\textit{data protection}'' and ``\textit{other security concerns}'' (P24), ``\textit{leaks}'' (P27), and the fact that other entities have ``\textit{the ability to steal data}'' (P28). P55 noted challenges with handling and securing data in ``\textit{central databases, where that data may be relied on by many loosely connected applications and systems}''. These responses highlight the difficulties of implementing mechanisms to safeguard users' data.

\myparagraph{Vague Requirements:} 10 survey respondents highlighted a lack of clear requirements as the biggest challenge with GDPR compliance in OSS. For example, one participant mentioned that GDPR ``\textit{is pretty vague}'' with a lack of ``\textit{standard format}'' (P54). Another described confusion in knowing ``\textit{how long can data be retained}'' and ``\textit{what is Personaly[sic] Identifiable Information}''---adding, the ``\textit{lack of clarity in the rgulations[sic] leads to confusion}'' (P52). Moreover, P48 highlighted the lack of company understanding of GDPR requirements makes compliance difficult.

Beyond these clear categories, we also received a wide range of other responses, including ``\textit{lawfulness and dispute resolution}'' (P47), the conflict between ``\textit{individual privacy and the public's right to know}'' (P21), and being in a ``\textit{rush to regulate}'' (P28). P27 mentioned challenges with user experiences, stating ``\textit{users endure invasive pop-ups}''. Further, P1 noted the challenges evolve during the lifetime of a project, stating ``\textit{At the beginning of a project, privacy by design and default. In the middle or the end, data minimization and transparency}'' are the main challenges.
Based on the challenges of implementation, participants described difficulties limiting functionality---\eg ``\textit{knowing when interacting with EU citizens}'' (P49) and ``\textit{more than 1,000 news websites in the European Union have gone dark}'' (P15). Meanwhile, P17 mentioned difficulties implementing GDPR requirements for data-intensive programming domains: ``\textit{many of the GDPR's requirements are essentially incompatible with big data, artificial intelligence, blockchain, and machine learning}''. These challenges motivate new resources to help developers overcome problems related to GDPR implementation and compliance.

\begin{tcolorbox} [width=\linewidth, colback=yellow!30!white, top=1pt, bottom=1pt, left=2pt, right=2pt]
\textbf{Finding 6:} The management and protection of user data and vague requirements are key challenges open-source developers face when implementing GDPR requirements.

\end{tcolorbox} 

\subsection{RQ4: Compliance Assessment}\label{sec:rq4}



This question was: \textit{RQ4: How do OSS developers assess GDPR compliance?}
We found three kinds of responses related to compliance assessment:
  consulting with legal counsel,
  referencing other compliance resources,
  and
  self-assessment.

\myparagraph{Compliance Through Legal Counsel:} In our survey results, 15 OSS developers reported consulting with legal teams for GDPR compliance. We were also interested in exploring the impact of seeking legal counsel for GDPR compliance on OSS development processes. Seven participants with experience seeking legal consultations noted that it did have a positive impact on development activity (P6, P13, P14, P45, P53, P55, P56). Participants noted the benefits of seeking legal experts, stating the importance of ``\textit{consulting with lawyers on the team who have a seat at the table}'' (P45), it ``\textit{clarifies requirements and prevents misinterpretations}'' (P55), and allowed GDPR compliance to be ``\textit{implemented rather easily}'' (P56). 

However, most participants ($n = 9$) with experience seeking legal counsel lamented the impact, stating it decreased development productivity: ``\textit{it slows things down as code has to be reviewed and objectives revised}'' and ``\textit{it impacted our approach to the SDLC}'' (P1), ``\textit{it's a bit of a headache}'' (P24), ``\textit{it slowed us down...was mostly a box ticking exercise}'' (P51), and ``\textit{it interrupted the development but it is required}'' (P49). Respondents also bemoaned the costs of working with legal teams, stating ``\textit{for a global project open source project any legal advice would be extremely expensive}'' (P52) and ``\textit{open-source projects can't afford even to sustain maintainers, not even speaking about legal team...Legal teams are consulted with some corps want to kill the project}'' (P47). P54 also noted legal experts found difficulties with the vagueness of GDPR compliance, replying the ``\textit{legal team struggles to interpret how to comply with GDPR, there are a lot of back-and-forth. We have to change our design many times}''.

In sum, legal experts can provide valuable insight into data privacy regulations and compliance, but developers often find these interactions negatively impact development processes.  

\myparagraph{Compliance Resources:} To assess GDPR compliance, three participants mentioned a variety of other resources. One participant described formal training on regulatory compliance, with a ``\textit{special training on GDPR within the company}'' (P16).  Another participant responded that their team uses an ``\textit{accountability system}'' (P24) to assess compliance.  Finally, P15 noted using online resources to help, but highlighted their ineffectiveness, stating, ``\textit{many of the articles on the Internet about GDPR are incomplete or even wrong''}.

\myparagraph{Self-assessment:} Other developers mentioned they were largely responsible for evaluating the ``\textit{legality}'' (P18) and ``\textit{integrity and confidentiality}'' (P23) of the processing and storage of user data in their system on their own. P24 responded developers have to ``\textit{consider whether you really need all the data you collect}'' while P38 advised to ``\textit{get your consent in order}''. P53 noted the impact on development teams, stating GDPR implementations ``\textit{took us significant amount of time due to several rounds of architecture review}''. 
P18 added there is ``\textit{really no good way}'' to evaluate compliance.


\begin{tcolorbox} [width=\linewidth, colback=yellow!30!white, top=1pt, bottom=1pt, left=2pt, right=2pt]
\textbf{Finding 7:} Developers often do not consult legal experts to validate GDPR compliance, relying on other resources such as compliance training, accountability systems, online resources, and self-assessed data management.

\textbf{Finding 8:} Participants with experience interacting with legal teams provided mixed perceptions, feeling they provided valuable insight but hindered development processes. 

\end{tcolorbox}

\section{Discussion and Future Work}

Our results demonstrate that GDPR-related code changes have a major impact in OSS development, significantly increasing development activity with regards to number of lines of code added and the number of commits included in PRs--indicating increased effort in code contributions and code review activities for developers (\cref{sec:rq1_prs}). Further, we found that GDPR compliance provides a wide range of challenges for OSS development (\cref{sec:rq3}) and that developers often assess compliance without the help of legal and policy experts (\cref{sec:rq4}). These findings posit that implementing GDPR compliance is a challenging activity for OSS developers. 

We recognize many stakeholders are involved in adhering to data privacy legislation. For instance, policymakers also play a role in data privacy compliance~\cite{weaver2015getting}. Data privacy regulations, such as the GDPR, are beneficial for protecting the rights and data of users online. However,  we noticed developers complaining about providing privacy to people--holding negative perceptions of the GDPR policy in general and its implementation. To that end, we provide guidelines to enhance data privacy regulations and software development processes to reduce the negative effects of policy compliance in OSS software.

\subsection{Improving Data Privacy Regulations}

\subsubsection{Provide Clear Requirements}

We found developers struggled to implement GDPR concepts (\cref{sec:rq3}). Moreover, few respondents reported consulting with legal experts to provide insight of policies and assess the compliance of projects (\cref{sec:rq4}). Thus, most development teams are forced to evaluate the system themselves. Yet, participants complained that understanding compliance is difficult due to the ambiguity of GDPR concepts: for instance, ``\textit{the procedure for obtaining user consent and the information provided are unclear}'' (P25). Prior work suggests ambiguity is a main challenge in requirements engineering~\cite{bano2015addressing}. Further, incomplete requirements can increase development costs and the probability of project failure~\cite{clancy1995chaos}. 

To improve program specifications, researchers have explored a variety of techniques. For instance, Wang \etal explored using natural language processing to automatically detect ambiguous terminology in software requirements~\cite{wang2013automatic}. Similar techniques could be applied to regulations such as the GDPR to notify policymakers of unclear language and clarify requirements for software engineers. Another way to improve the clarity of requirements is to involve software developers in the policy-making process. Verdon argues a good policy must be ``understandable to [its] audience''~\cite[p.~48]{verdon2006security}, yet our results show developers are confused by GDPR requirements. Prior work shows collaboration between policy makers and practitioners improves policies in domains such as public health~\cite{choi2005can} and education~\cite{ion2019can}. Thus, developers should be incorporated into the policy-making process to provide input on the impact of implementing and complying with policies concerning software development, such as data privacy regulations. 


\subsubsection{Policy Resources}

Our survey results show OSS developers face challenges implementing GDPR-related changes (\cref{sec:rq3}). Participants also found legal consultations negatively affect development processes (\cref{sec:rq4}), and report existing resources are largely ineffective, primarily relying on self-assessment within the development team. Only one participant mentioned receiving formal training on GDPR compliance (P16). To that end, OSS developers largely resort to implementing and evaluating compliance on their own efforts with ``\textit{insufficient information}'' (P26).  Prior work also outlines issues with software developers and security policies, noting a lack of understanding from programmers~\cite{verdon2006security}. 

 
 Based on our findings, we posit OSS development can benefit from novel resources to educate developers on policies and their implementation. To further support compliance, policymakers can provide resources, such as guides or online forums, to provide information on data privacy-related concepts in an accessible manner. These guidelines can also reduce the effects of GDPR compliance on code review tasks by providing specialized expertise and correct understanding for reviewers~\cite{pascarella2018information}. Yet, there are limited online developer communities focused on seeking help in data privacy policy implementation. Popular programming-related Q\&A websites, \eg Stack Overflow, are frequently used by developers to ask questions and seek information online~\cite{peterson2019gaze}---and are used for discussions on data privacy policy implementation (see Table~\ref{table:DataAvailabilityByPrivacyLegislation}). However developers have no way to verify the correctness of responses, which can also become obsolete over time. Zhang \etal recommend automated tools to identify outdated information in responses for development concepts, such as API libraries and programming languages~\cite{zhang2019empirical}. A similar approach can be used to keep responses regarding GDPR compliance up-to-date and accurate. 

\subsection{Improving Development Processes}

\subsubsection{Privacy by Design}
   
Participants reported challenges implementing GDPR compliance (\cref{sec:rq3}) and negative effects on development practices (\cref{sec:survey}). Moreover, our GitHub analysis found GDPR-related changes necessitated significantly more time and effort (\ie comments, commits, etc.) for developers to implement and review in PRs (see Table~\ref{tab:rq1_metrics}). However, compliance is required for organizations to avoid ``\textit{paying out hefty fines}'' (P30). Researchers have investigated techniques to streamline the incorporation of privacy in development processes. For instance, Privacy By Design (PBD) is a software development approach to make privacy the ``default mode of operation''~\cite{cavoukian2009privacy}. P50 mentioned cultivating ``\textit{a privacy-respecting mindset long before GDPR came about}'' avoided negative impacts on development processes and made the effort required ``quite minimal''. However, numerous participants noted the burden of implementing GDPR requirements, with one survey participant in particular (P1) highlighting that prioritizing privacy in software development processes ``\textit{requires an overhaul}''. Additionally, while PBD can benefit GDPR compliance efforts, Kurtz \etal note a scarcity of research in this area and note particular challenges with PBD for GDPR implementations, such as ensuring third party libraries also adhere to privacy principles~\cite{kurtz2018privacy}. 


PBD can be effective for new projects starting from scratch~\cite{tamo2018privacy}, yet may be ill-equipped for existing projects complying with new and changing data privacy regulations. Anthonysamy \etal outline limitations with current privacy requirements that solve present issues, which may differ from regulations and policies in the future~\cite{anthonysamy2017privacy}. More work is needed to explore tools and processes to support data privacy in mature software projects. One solution could be a partial or gradual approach to compliance. For instance, some programming languages (\eg Typescript) support gradual typing to selectively check for type errors in code~\cite{siek2007gradual}. Similarly, research in formal methods has explored supporting gradual verification of programs~\cite{bader2018gradual}. Thus, gradually introducing privacy into OSS can help reduce efforts related to GDPR compliance as opposed to overhauling development processes to prioritize privacy.


\subsubsection{Automated Tools}
We found GDPR compliance has a major impact on OSS development, significantly increasing coding and reviewing tasks for PRs in GitHub repositories  (see Table~\ref{tab:rq1_metrics}). Developers who responded to our survey also indicated the impact of GDPR compliance on their project source code, noting data privacy regulations always need more software (P4) and violate the principle of minimum scope (P21). This indicates further difficulty for developers to validate their projects for the GDPR, with one participant responding there is ``\textit{no good way}'' to assess compliance (P18). These findings point to an increased burden and effort on OSS developers to implement and review GDPR requirements to comply with data privacy regulations and avoid penalties for non-compliance (\eg losing market share).

To that end, we posit automated tools can reduce the burden of GDPR implementation efforts. One participant mentioned using a tool, an ``\textit{accountability system}'' (P24), to help assess compliance--however did not provide any details about this system. Our findings for RQ1 (\cref{sec:rq1}) show GDPR-related pull requests have significantly more coding involved, consisting of more commits and lines of code added in code contributions, as well as requiring significantly more comments and time in reviewing processes. Thus, systems to support data privacy implementation and tools to review policy-relevant code are needed to streamline compliance. Ferrara and colleagues present static analysis techniques to support GDPR compliance~\cite{ferrara2018static}. Further tools can support review processes for assessing implementation changes. Prior work suggests static analysis tools can reduce time and effort in code reviews~\cite{singh2017evaluating}. Future systems could also provide automated feedback to developers and reviewers on data privacy regulation compliance. For instance, using NLP techniques~\cite{aberkane2021exploring} or rule-based machine learning approaches~\cite{hamdani2021combined} to automatically summarize requirements and verify compliance. 


\subsection{Other Directions}

Based on our results, we observe several other avenues of future work. First, we plan to investigate other data sources to further explore GDPR compliance in open-source projects. For example, we plan to mine relevant queries from Stack Overflow to gain insight into challenges and information needs developers have for implementing GDPR policies. We will also examine answers to observe how developers respond. For instance, online discussions between developers regarding policies often use disclaimers, such as the acronyms ``IANAL'' or ``NAL'' to indicate ``I am not a lawyer'', 
before offering advice or answering questions related to legal frameworks. Without legal expertise, we anticipate it is difficult for OSS developers to offer guidance and seek help complying with data privacy regulations--motivating the need for novel approaches to support regulation adherence and compliance assessment. 

Moreover, we aim to engage with policymakers to understand their perspectives on data privacy policies and the challenges developers face implementing them. We will collect qualitative insights from politicians and individuals with authority to develop policies to further explore methods to support the implementation of privacy laws. Finally, we aim to extend this work to investigate the impact of broader technology-related policies on open-source software development practices--for instance, investigating the impact of alternative data privacy regulations (\ie the CCPA or CDPA) as well as other legal frameworks that will impact software development and maintenance, such as current and imminent legislation regarding artificial intelligence governance.

\section{Related Work} 

We note two lines of related work: characterizations of stakeholder perspectives on data privacy regulations, and technical and methodological approaches for regulatory compliance.

\myparagraph{Stakeholder perspectives:}
    Research has investigated perspectives on the GDPR for stakeholders in data privacy regulation compliance. Sirur and colleagues examined organizational perceptions on the feasibility of implementing GDPR concepts, finding that larger organizations were confident in their ability to comply while smaller companies struggled with the breadth and ambiguity in GDPR requirements~\cite{10.1145/3267357.3267368}. Earp \etal surveyed software users to show the Internet privacy protection goals and policies for online websites do not meet users' expectations for privacy~\cite{1424412}. Similarly, Strycharz \etal surveyed consumers to uncover frustrations and negative attitudes related to the GDPR~\cite{strycharz2020data}. Our work focuses on the perceptions of developers, who are responsible for implementing code changes to comply with data privacy regulations.

    On the perspective of software engineers as regulatory stakeholders, 
    van Dijk and colleagues provide an overview of the transition of privacy policies from self-imposed guidelines from developers to legal frameworks and legislation~\cite{van_Dijk_Tanas_Rommetveit_Raab_2018}. Alhazmi interviewed software developers to uncover barriers for adopting GDPR principles--finding the lack of familiarity, precedented techniques, useful help resources, and prioritization from employers. The paper also found that developers generally do not prioritize privacy features in their projects, focusing instead on functional requirements prevent compliance~\cite{alhazmi_arachchilage_2021}. Similarly, researchers interviewed senior engineers to understand the challenges implementing general privacy guidelines, indicating a frustration with legal interactions and the non-technical aspects of requirements~\cite{doi:10.1080/01972243.2019.1583296}. Finally, Klymenko \etal interviewed technical and legal professionals to investigate measures for data privacy compliance in GDPR implementation---noting a lack of understanding and need for interdisciplinary solutions~\cite{klymenko2022understanding}.
    While these papers take similar approaches to our research, ultimately our goals and questions are distinct, since we are specifically interested in the perspective of open-source developers.
    
\myparagraph{Implementing and verifying GDPR compliance:}
    Prior work has explored approaches to implement and verify GDPR compliance. For instance, Mart\'{i}n \etal recommend Privacy by Design methods and tools for GDPR compliance~\cite{martin2018methods}. Shastri and colleagues introduce GDPRBench, a tool to assess the GDPR compliance of databases~\cite{shastri_banakar_wasserman_kumar_chidambaram_2020}. Li \etal investigated automated GDPR compliance as part of continuous integration workflows~\cite{8933680}.
    Al-Slais conducted a literature review to develop a taxonomy privacy implementation approaches to guide GDPR compliance~\cite{9311949}.
    Finally, Mahindrakar \etal proposed the use of blockchain technologies to validate personal data compliance~\cite{9123033}.
    Rather than proposing new software engineering methods, measures, and tools related to GDPR, our work takes an empirical perspective to understand current practices.

\section{Threats to Validity}


We discuss three types of threats to validity.

\myparagraph{Construct:}
In mining OSS repositories, we defined the construct of ``GDPR-related pull requests'' based on the presence of the string ``GDPR''.
Some PRs may incorrectly refer to GDPR (false positives), while others may perform GDPR-relevant changes without using the acronym (false negatives). This is also biased towards English-speakers, as this acronym differs in other languages. To mitigate non-English GDPR-related PRs polluting the non-GDPR-related dataset, we manually inspected PR titles for various iterations of the GDPR in other languages, including ``RGPD'' (French, Spanish, and Italian), ``DSGVO'' (German), and ``AVG'' (Dutch). However, these were not included in our GDPR-related dataset since we only focus on PRs in English for our analysis. We used off-the-shelf NLP techniques to assess sentiment, inheriting biases from these methods (\eg misinterpreted connotations of homonyms such as ``mock''). In addition, parametric models for sentiment analysis are based on defined dictionary values and cannot detect certain aspects of human communication, such as sarcasm. Prior work also suggests sentiment analysis tools can be inaccurate in software engineering contexts~\cite{jongeling2017negative}. However, we use this to gain preliminary insights into developers' perceptions of GDPR compliance in OSS.

\myparagraph{Internal:}
We perceive no internal threats.
This study provides characterizations rather than cause-effect measurements.

\myparagraph{External:}
There are several threats to the generalizability of our findings.
We inherit the standard perils of mining open-source software~\cite{kalliamvakou2014promises}.
We focus on open-source software available on GitHub, which omits other code hosting platforms, such as GitLab, which may be used by different populations of developers.
We doubt our results generalize to commercial software, since those development organizations directly face the consequences of GDPR non-compliance.
We only consider the effect of GDPR because it is the most prominent privacy law, and hence has the most available data.
Other regulations may have different effects.
Specifically, we conjecture differences in the software engineering impact between general data privacy regulations, such as the GDPR and CCPA, and industry-specific data privacy regulations, such as HIPAA and FERPA: general regulations may necessarily be more ambiguous.


\section{Conclusions}
Data privacy regulations are being introduced to prevent data controllers from misusing users' information and to protect individuals. To adhere with these regulations, developers are charged with the complex task of understanding policies and making modifications to the source code of applications to implement privacy-related requirements. This work examines the impact of data privacy regulations on software development processes by investigating code contributions and developer perceptions of GDPR compliance in open-source software. Our results show that complying with data privacy regulations significantly impacts development activities on GitHub, evoking negative perceptions and frustrations from developers. Our findings provide implications for developers and policymakers to support the implementation of data privacy regulations that protect the rights of human users in digital environments.



\section{Data Availability}

We have uploaded the survey, datasets, and data collection and analysis scripts as supplementary materials~\cite{data}.
Our IRB protocol does not allow us to share individual survey responses. 


\section{Acknowledgments}

Brown and Brantly acknowledge support from the Virginia Commonwealth Cyber Initiative (CCI).

\fi

\balance
\newpage

\bibliographystyle{ACM-Reference-Format}
\bibliography{davis,master}

\end{document}